\documentstyle[preprint,aps]{revtex}
\input epsf.tex
\def\DESepsf(#1 width #2){\epsfxsize=#2 \epsfbox{#1}}
\begin{document}

\preprint{\vbox{\hbox{OITS-638}\hbox{OSURN-330}}}
\draft
\title {An Unusual Signal for Supersymmetry at the Tevatron }
\author{B. Dutta$^{1} $ and S. Nandi$^{2}$}
\address{$^{1} $ Institute of  Theoretical Science, University of Oregon,
Eugene, OR 97403\\$^{2} $ Department of Physics, Oklahoma State University,
Stillwater, OK 74078}
\date{October, 1997}
\maketitle
\begin{abstract} 

We propose a new scenario in which the dominant signal for supersymmetry at
the Tevatron are the events having two or three $\tau$ leptons  with
high
$p_T$ accompanied by  large missing transverse energy. This signal is very
different from the multijet or multileptons (involving $e$ and/or $\mu$ only) or
the photonic signals that have been extensively investigated both theoretically
and experimentally. A large region of the GMSB parameter space with the lighter
stau as the NLSP allow this possibility. Such a signal may be present in the past
Tevatron data to be analyzed.

\end{abstract}
\newpage

\newpage 
 Over the past two decades, there has been a great deal of experimental effort
to  discover the supersymmetric (SUSY) particles at the Tevatron. The searches
have concentrated mainly in two fronts. One is to look for the SUSY partners of
the strongly interacting quarks and gluons, namely the squarks and the
gluinos\cite{cdfD01}. In this case, the signals are multijets and/or
multileptons with high $p_T$ plus large missing transverse energy
(${\rlap/E}_T$). The other is to look for the SUSY partners of the leptons and
the electroweak (EW) gauge bosons namely the sleptons, charginos and the
neutralinos. The signals looked for in this case are high
$p_T$ multileptons, namely dileptons and trileptons (involving e and/or $\mu$) 
\cite{cdfD02} accomapanied by large ${\rlap/E}_T$. Both types of the
signals have been motivated by the gravity mediated SUSY breaking theories where
the lightest neutralino is  mostly the lightest SUSY particle (LSP) to which
 all the produced SUSY particles eventually decay. LSP escapes detectors giving
rise to the ${\rlap/E}_T$ signal. In the last two years gauge mediated SUSY
breaking (GMSB) models have become very popular \cite{{DN},{DMN}}. In these
theories, gravitino is the LSP, and most of the experimental searches have
concentrated on the regions of the theory in which the lightest 
neutralino $\chi_0$ is the next to lightest SUSY particle (NLSP). $\chi_0$
decays to a photon and a gravitino within the detector. The signal in this case
is photons with high
$p_T$ and large ${\rlap/E}_T$ \cite{DWR}. Inspite of the extensive searches no
SUSY signal in any of the considered  modes has yet been observed either at the
Tevatron
\cite{cdfD03} or at the LEP2
\cite{LEP}, except for one possible
$e^+e^-\gamma\gamma$ plus missing energy event at the Tevatron \cite{SP}. The
object of this work is point out that in the GMSB theories the dominant signal
for the SUSY can be two to three high $p_T$ $\tau$ leptons plus large
${\rlap/E}_T$. This happens when the lighter of the scalar tau ($\tilde \tau_1$)
is the NLSP. In that case all the produced SUSY particles will eventually decay
to
$\tilde\tau_1$ which subsequently decays to
a $\tau$ and a gravitino. The gravitino escapes the detector giving rise to
${\rlap/E}_T$. Such a signal occurs for a large region of the GMSB parameter
space. In this case, discovery of SUSY will crucially depend on the ability of
the detectors to detect high $p_T$ $\tau$ leptons with good efficiency.

The dominant processes that give rise to our signal at the Tevatron are the
productions of chargino pairs ($\chi^+\chi^-$) and the chargino and the second
neutralino pairs ($\chi^\pm\chi^0_{2}$). ($\sigma_{\chi^{\pm}\chi^0}$, where
$\chi_0$ is the lightest neutralino is very small compared to
$\sigma_{\chi^{\pm}\chi^0_2}$). The chain of decays of
$\chi^\pm$ and $\chi^0_{2}$ lead to the observable high $p_T$ $\tau$'s and the
missing neutrinos and the gravitinos. There are several mass hierarchies of these
superparticles, (involving $\chi^\pm, \,\chi^0_{2,}\,\chi^{0},\,\tilde\nu_l
(l=e,\mu,\tau),\,\tilde l (l=e,\mu \, {\rm and}\, \tilde\tau_1$) leading to the
inclusive high $p_T$ 2$\tau$ or 3$\tau$ final states plus ${\rlap/E}_T$. Below we
first briefly discuss the GMSB parameter space  giving rise to $\tilde \tau_1$
as the NLSP and our desired scenario.

In GMSB models, with radiative EW symmetry breaking, all the sparticle masses
and the mixing angles depend on five parameters,
$M,\,\Lambda,\,n,\,\tan\beta,\,{\rm and\, sign\, of\,}
\mu$.
$M$
 is the messenger scale, and $\Lambda$ is equal to $\langle F_s\rangle/\langle
s\rangle$, where
$\langle s\rangle$ is the VEV of the hidden sector field $S$, and 
$\langle F_s\rangle$ is the VEV of the auxillary component of $S$.  The parameter
n is fixed by the choice of the vector like messenger sector. For example, for 
$5+{\bar 5}$ of
$SU(5)$, n  can take the values 1,2,3 or 4.  The parameter $\tan\beta$ is the
usual ratio of the up ($H_u$) and down ($H_d$) type Higgs VEVs. The parameter
$\mu $ is the coefficient in the bilinear term, $\mu H_uH_d$ in the
superpotential.  We demand that the electroweak symmtery be broken radiatively
which then determines the magnitude of $\mu$ and another parameter 
$B$ (in $B\mu H_uH_d$ term in the potential) in terms of the other parameters of
the theory. The constarints coming from
$b\rightarrow s\gamma$ demands $\mu$ to be mostly negative \cite{ddo}. The soft
SUSY breaking gaugino and the scalar masses at the messenger scale M are given by
\cite{{DN},{SPM}}
\begin{equation}
\tilde M_i(M) = n\,g\left({\Lambda\over M}\right)\,
{\alpha_i(M)\over4\pi}\,\Lambda.
\end{equation} and
\begin{equation}
\tilde m^2(M) = 2 \,(n)\, f\left({\Lambda\over M}\right)\,
\sum_{i=1}^3\, k_i \, C_i\,
\biggl({\alpha_i(M)\over4\pi}\biggr)^2\,
\Lambda^2.
\end{equation} where $\alpha_i$, $i=1,2,3$ are the three SM gauge couplings and
$k_i=1,1,3/5$ for SU(3), SU(2), and U(1), respectively. The $C_i$ are zero for
gauge singlets, and 4/3, 3/4, and $(Y/2)^2$ for the fundamental representations
of
$SU(3)$ and $SU(2)$ and
$U(1)_Y$ respectively (with Y defined by $Q=I_3+Y/2)$. Here $n$ corresponds to
$n(5+{\bar 5})$. $g(x)$ and
$f(x)$ are messenger scale threshold functions with $x=\Lambda/M$.  We have
studied in great detail the super partner mass spectrum in the $(M,
\Lambda, n, \tan\beta)$ parameter space for negative $\mu$. We have calculated
the SUSY mass spectrum using the appropriate RGE equations
\cite{BBO} with the boundary conditions given by the equations above. In the
scenarios where 
$\chi_0$ is the NLSP, one obtains high $p_T$ photon signal and these scenarios
have been studied in great detail both theoretically 
\cite{{dwt},{bkw}}and
experimentally \cite{{cdfD03},{LEP}}. In the GMSB parameter space, this occurs
for
$n=1$ and low values $\tan\beta$ ($\tan\beta\le25$) \cite{dwt}. However, as
$\tan\beta$ increases,
$\tilde\tau_1$ becomes the NLSP for most of the parameter space with lower
values of
$\Lambda$. For $n\ge2$, $\tilde\tau_1$ is the NLSP even for the low values of
$\tan\beta$ (for example, $\tan\beta\gtrsim 2$), and for $n\ge3, $
$\tilde\tau_1 $ is again naturally the NLSP for most of the parameter space. The
parameter space where $\tilde\tau_1$ is the NLSP  give rise to our proposed high
$p_T$ $\tau$ signals. The next question we ask is the following : can
$\chi^{\pm}$ and
$\chi^{0}_{2}$ be light enough to be produced at the Tevatron energy? Some
examples of the mass spectrum, appropriate for exploration at  the Tevatron
enrgy is shown in Table 1 and 2. All of the superpartner masses satisfy the
current experimental limits applicable for this scenario.
 
At the Tevatron, the chargino pair ($\chi^+\chi^-$) production takes place
through the s-channel  Z and $\gamma$ exchange; while the 
$\chi^0_{2}\chi^{\pm}$ production is via the s channel W exchange. Squark
exchange via the t-channel will also contribute to both  processes. Since the
squark masses are large in the GMSB models these contributions are negligible.
The inclusive final states arising from both of processes are either 2$\tau$ or
3$\tau$ with high
$p_T$ plus ${\rlap/E}_T$ (due to the undetected neutrinos and gravitinos). The
2$\tau$ mode will be mostly of opposite sign charges, but a significant
fraction will also have the same sign of charges. The details of the decays for
the
$\chi^{\pm}$ and $\chi^0_{2}$ depend on the hierarchies of the superparticle
masses, which in turn depend on the GMSB parameter space. In most of the
parameter space we looked at, the mass of $\chi^0_{2}$ is approximately equal to
that of 
$\chi^{\pm}$ (within a few GeV). The sneutrinos $\tilde \nu_l$
(l=e,\,$\mu$,\,$\tau$) can be heavier or lighter than the
$\chi^{\pm}$. (The sneutrinos are always heavier than the righthanded sleptons
because of the SU(2) gauge interactions.) Also,
$\chi_0$ can be heavier or lighter than the right handed $\tilde e,\,
\tilde\mu$. Thus there are four possible cases for the mass hierarchies:
\begin{eqnarray} {\rm case 1}&:& M_{\chi^0_{2}}\geq M_{\chi^\pm}>m_{\tilde
\nu}>m_{\tilde {e},\tilde {\mu}}>M_{\chi^{0}}>m_{\tilde \tau_1}\\\nonumber {\rm
case 2}&:&M_{\chi^0_{2}}\geq M_{\chi^\pm} >m_{\tilde \nu}>M_{\chi^{0}}>m_{\tilde
{e},\tilde {\mu}}>m_{\tilde \tau_1}\\\nonumber {\rm case 3}&:& m_{\tilde \nu}>
M_{\chi^0_{2}}\geq M_{\chi^\pm}>m_{\tilde {e},\tilde
{\mu}}>M_{\chi^{0}}>m_{\tilde
\tau_1}\\\nonumber {\rm case 4}&:& m_{\tilde \nu}> M_{\chi^0_{2}}\geq
M_{\chi^\pm}>M_{\chi^{0}}>m_{\tilde {e},\tilde {\mu}}>m_{\tilde
\tau_1}\\\nonumber
\end{eqnarray} All three sneutrino masses are essentially the same. The scalars
$\tilde e$ and $\tilde
\mu$ are essentially right handed and get the same masses. In table 1, we give
four scenarios for case 1. The scenarios occur mostly for n=3. The reason is that
the gaugino masses are proportional to n while the scalar masses are proportional
to
$\sqrt n$. Thus for higher values of n, the scalar masses get reduced compared to
the gaugino masses resulting in $m_{\tilde\nu}$,
$m_{\tilde e,\tilde \mu}<M_{\chi^{\pm}}$. Increasing n further(n=4), $\tilde e$,
$\tilde \mu$ masses get even smaller than $m_{\chi_0}$ producing case 2. Two
scenarios for case 2 is shown in Table 2. Case 3 occurs for smaller values of n
in order to have $m_{\tilde\nu}$ larger than $M_{\chi^0_{2}}$. Two scenarios for
case 3 are also shown in table 2. We have not found a scenario for case 4. In
table 1 and 2, we also give the production cross-sections of the chargino pair
and for the chargino -second neutralino at the Tevatron for two center of mass
energies,
$\sqrt s=1.8 $ TeV and $\sqrt s$=2.0 TeV. $\sigma (\chi^{\pm}\chi^0_{2})$ is
always somewhat bigger than $\sigma (\chi^+\chi^-)$.

Now, we are ready to discuss the decays of the chargino and the second
neutralino and the resulting final states. We first consider the case 1. The
four main decay modes for the chargino are:$\chi^+\rightarrow
(\nu_\tau\tilde\tau_1,\,\tau\tilde\nu_\tau,\, e\tilde\nu_e,\,\mu\tilde
\nu_\mu)$. (Since the lighter $\tilde e$ and $\tilde \mu$ are essentially right
handed the branching ratio to the other two kinematically allowed decay modes
$\nu_e\tilde e$, $\nu_\mu\tilde \mu$ are essentially zero.)  Subsequently
$\tilde \nu_l (l=e,\,\mu,\,\tau)$ decays to $\nu_l\chi^{0}$,
followed by the decay
$\chi^{0}\rightarrow
\tau\tilde\tau_1$. Finally, $\tilde\tau_1$ decays to a $\tau$ and a gravitino.
(The branching ratio for the deacys of $\tilde\nu_l$, $\tilde l$, $\chi^{0}$ and
$\tilde\tau_1$ is 100 $\%$.) The possible final states from a chargino decays
are :
$(\tau,\, 3\tau,\, e\tau\tau, \,\mu\tau\tau) $ accompanied by the neutrinos and
the gravitinos. Only one
$\tau$ arising from the decay of $\tilde\tau_1\rightarrow\tau\tilde G$ will have
large $p_T$. The final states arising from the decays of the produced chargino
pair are: $2\tau$, 4
$\tau$,
$6\tau$, $(e, \mu) + 3 \tau$, $(e, \mu) + 5 \tau$ ,$(2e, 2\mu) + 4 \tau$,
$e\mu + 4 \tau$, accompanied by the large ${\rlap/E}_T$ due to the undetected
gravitinos and the neutrinos. The branching ratios for these various final
states for the scenarios in case 1 are  given in table 3. Note that among the
multileptons, only the two $\tau$'s coming from the decays of the
$\tilde\tau_1$'s will have large
$p_T$. This happens for all the final states. Thus, the  inclusive final state
to look for is two high $p_T$ $\tau$ s accompanied by large
${\rlap/E}_T$. These two high $p_T$ $\tau$'s can have both opposite as well as
same sign of charges. For the large number of scenarios we looked at, the  ratio
$(\tau^-\tau^-+\tau^+\tau^+)/(\tau^+\tau^-)$ lies between 0.5 to 0.7. Note that
the values of $\sigma.B$, given in the upper half of Table 3, are for the final
states having 2 high $p_T$ $\tau$'s accompanied by any number of soft charged
leptons( $e,\,\mu$ or $\tau$). The inclusive
cross-sections are given in table 1 and can be as large as almost half a
picobarn at $\sqrt s=$ 1.8 TeV at the Tevatron.

Next we discuss the decays of the second lightest neutralino, and the resulting
final states when it is produced in association with a chargino. Again, we first
consider case 1. The six decay modes of second neutralino are 
$\chi^0_{2}\rightarrow \tau\tilde\tau_1,\,e\tilde e,\, \mu\tilde
\mu,\,\nu_{\tau}\tilde\nu_{\tau},\,\nu_{\mu}\tilde\nu_{\mu},\,\nu_{e}\tilde\nu_{e}$.
Subsequently, $\tilde l$ (l=$e$, $\mu$) decays to $l\chi^{0}$, and $\chi^{0}$,
$\tilde\nu_l$ and $\tilde\tau_1 $ decay as before. Thus the possible final
states from a $\chi^0_{2}$ decay are 
$(\tau\tau,\, ee\tau\tau, \, \mu\mu\tau\tau)$ accompanied by the neutrinos and
gravitinos. The final states resulting from the
$\chi^0_{2}\chi^{\pm}$ productions and their subsequent decays are:$3\tau$, 5
$\tau$,
 $(e, \mu) + 4 \tau$, $(2e, 2\mu) + 3 \tau$  $(2e, 2\mu) + 5
\tau$, $(2e\mu, 2\mu e) + 4 \tau$, $(3 e, 3 \mu) + 4 \tau$, accompanied by the
large ${\rlap/E}_T$ due to the undetected gravitinos and the neutrinos. The
branching ratios for the various final states for the scenarios in case 1 are
given in table 3. Note that here we can
have events with 2$\tau
$' s having high
$p_T$ or events with 3 $\tau$'s having high $p_T$. The latter happens when
$\chi^0_{2}$ decays into $\tau$ and
$\tilde\tau_1$. The ratio of the 2$\tau
$ vs 3$\tau$ events are scenario dependent as can be seen from the values of
$\sigma.B$ given in the lower part in the table 3. In the case of high
$p_T$ 2$\tau$ events, we have equal number with the same or opposite sign
charges. Again, note that the values for $\sigma.B$, given in the lower half of
table 3, are for the final states having two or three high $p_T$ $\tau$'s
accompanied by any number of soft charged leptons. The inclusive cross-sections
are given in table 1 and can be more than half a picobarn at
$\sqrt s=$ 1.8 TeV at the Tevatron.

We now briefly discuss case 2 and case 3. In table 2 we give two scenarios (5 and
6) for the case 2, and two scenarios (7 and 8) for the case 3. In case 2, there
are  two additional decay modes for
$\chi^{0}$ ($\chi^{0}\rightarrow e\tilde e,\,
\mu\tilde \mu$) giving rise to more overall final states for the $\chi^+$ and
$\chi^0_{2}$ decays. For the case 3, both $\chi^+$ and $\chi^0_{2}$ has less
decay channels (since $\tilde \nu_l$'s (l=e, $\mu$, $\tau$) are heavier), and
hence less final states. However, both cases give rise to inclusive 2$\tau$ and
3$\tau$ high $p_T$ final states (accompanied by the missing transverse energy)
with characteristics very similar to the case 1.

 What is the prospect of detecting this signal in the past Tevatron data, in
ongoing LEP2 experiments or in the upgraded Tevatron? CDF collaboration is
capable of detecting $\tau$ leptons via their jet decay modes or via their $e$
(or
$\mu$) decay modes. This requires high $p_T$ for each $\tau$ and large
${\rlap/E}_T$ in the events. In  our scenario, in the high $p_T$ 2$\tau$ signal,
both
$\tau$'s come from the decay $\tilde\tau_1\rightarrow\tau\tilde G$, while in the
high $p_T$ 3$\tau$ signal, 2$\tau$'s come from the above decay while the third 
$\tau$ comes from the decay $\chi^0_{2}\rightarrow\tau\tilde\tau_1$. Since both
$\tilde\tau_1$ and
$\chi^0_{2}$ are very heavy, these signal $\tau$ leptons are expected to be have
high
$p_T$. Since several neutrinos and the gravitinos escape detection in each
event, large ${\rlap/E}_T$ is expected even after cancellations. No dedicated
search for this signal has yet been carried out \cite {c}. With a total (2$\tau$
+3 $\tau$) high $p_T$ inclusive cross-section as large as 1 pb (see table 1
and 2), few signal events should be found in the CDF $110 \,pb^{-1}$ data, even
with an overall acceptance of few percent for these events. Upgraded Tevatron
will see such events for much heavier charginos and  second lightest
neutralinos. We give results for the cross-sections for the $\chi^+\chi^-$ and
$\chi^{\pm}\chi^{0}_2$ pair productions in fig 1. These cross-sections have been
calculated in a large region of the GMSB parameter space giving rise to our
scenarios. For a given chargino mass, the cross-sections are approximately
scenario independent. The dashed curves are for the case, $\sqrt s$=1.8 TeV,
while the solid curves are for the case $\sqrt s$=2.0 TeV. In each case, the
bottom ones are for
$\sigma(\chi^+\chi^-)$, the middle ones are for $\sigma(\chi^{\pm}\chi^0_2)$ and
the top ones are for the sum of the two. At LEP2, our scenarios will produce
4$\tau$ events with large missing energy from the production of a
$\chi_0$ pair and their subsequent decays \cite{we1}.  Plans
are underway to look for such a signal at LEP2\cite{gw}. 

At Tevatron, the backgrond for our high $p_T$ 2$\tau$ or 3$\tau$ signal events
could come from the WW, WZ, ZZ pair productions and Drell-Yan pairs, fake
$\tau$'s, W+jets. After appropriate cuts, the expected number of 2$\tau$ events
from these background is about two \cite{c}. The expected background for 3$\tau$
signal events is even much smaller. Charged Higgs or
$\tilde\tau_1$ pair productions can also give 2$\tau$ signal events, however
these cross-sections are very low at the Tevatron.

We conclude summarizing our main points. We have proposed a scenario in
which the dominant signal for supersymmetry at the Tevatron are the events with
two or three high $p_T$ $\tau$ leptons accomapnied by large missing
transverse energy. This happens in the GMSB models with $\tilde\tau_1$ as the
NLSP for a wide region of the parameter space. The signal events are jet-less.
There will be no high $p_T$ dileptons (involving electrons or muons) or photons
in this scenario. In the 2$\tau$ events, the $\tau$s can be of same sign as well
as of opposite sign. Our signal events may already be present in the past
Tevatron data. A dedicated ``2$\tau$", ``$3\tau$" search will be needed to
detect such a signal. Discovery of SUSY in such a scenario in the future
Tevatron run will crucially depend on the ability of the detectors to detect
high $p_T$ $\tau$ leptons with good efficiency. 

 We are grateful to John Conway, Stephan Lammel and David Stuart of the CDF
collaboration for many useful discussions regarding the $\tau$ detection and the
feasibility of observing the proposed events at the Tevatron. This work was
supported in part by the  US Department of Energy Grants No. DE-FG06-854ER-40224
and DE-FG02-94-ER40852.

\newpage
\begin{center}  Table 1 \end{center}
\begin{center}
\begin{tabular}{|c|c|c|c|c|}  \hline  &Scenario 1&Scenario 2&Scenario3&Scenario
4\\\hline  &$\Lambda=18$ TeV,&$\Lambda=20$TeV,&$\Lambda=60$ TeV,&$\Lambda=59.7$
TeV,\\ &n=3,$M= 40\Lambda$&n=3, $M=60\Lambda$&n=3, $M=20\Lambda$&n=3,
$M=40\Lambda$\\
&$\tan\beta$=18&$\tan\beta$=18&$\tan\beta$=19&$\tan\beta$=12\\\hline  m$_h
$(GeV)&111&113&113&109\\\hline m$_{H^{\pm}}$&224&242&268&216\\\hline 
m$_A$&210&229&256&201\\\hline m$_{\chi^0}$&67&75&81&61\\\hline 
m$_{\chi^0_2}$&114&129&142&106\\\hline
m$_{\chi^0_3}$&-202&-216&-243&-191\\\hline 
m$_{\chi^0_4}$&238&254&275&230\\\hline
m$_{\chi^{\pm}}$&112,240&127,255&140,276&101,231\\\hline 
m$_{\tilde{\tau}_{1,2}}$&51,141&57,151&57,162&60,130\\\hline 
m$_{\tilde{e}_{1,2}}$&72,132&77,143&81,153&70,125\\\hline
m$_{\tilde{\nu}}$&105&119&130&97\\\hline  m$_{\tilde {\rm
t}_{1,2}}$&384,465&427,506&439,528&365,446\\\hline  m$_{\tilde {\rm
b}_{1,2}}$&403,427&449,472&469,496&384,401\\\hline  m$_{\tilde {\rm
u}_{1,2}}$&417,430&463,478&488,504&394,405\\\hline  m$_{\tilde {\rm
d}_{1,2}}$&419,438&465,485&488,510&395,413\\\hline  m$_{\tilde
g}$&480&533&560&453\\\hline
$\mu$&-191&-206&-233&-179\\\hline
$\sigma_{p\bar{p}\rightarrow \chi^+\chi^-}$&427.43&250.55&172.42&635.20\\
(fb)&(508.26)&(304.45)&(213.85)&(744.34)\\\hline
$\sigma_{p\bar{p}\rightarrow \chi^{\pm}\chi^0_2}$&547.48&314.82&222.04&776.49\\
(fb)&(669.05)&(394.76)&(284.78)&(933.60)\\\hline 
\end{tabular}
\end{center}
\newpage
\begin{center}  Table 2 \end{center}
\begin{center}
\begin{tabular}{|c|c|c|c|c|}  \hline  &Scenario 5&Scenario 6&Scenario7&Scenario
8\\\hline  &$\Lambda=18$ TeV,&$\Lambda=15$TeV,&$\Lambda=28$ TeV,&$\Lambda=30$
TeV,\\ &n=4,$M= 40\Lambda$&n=4, $M=40\Lambda$&n=2, $M=15\Lambda$&n=2,
$M=4\Lambda$\\
&$\tan\beta$=18&$\tan\beta$=12&$\tan\beta$=20&$\tan\beta$=25\\\hline  m$_h
$(GeV)&115&111&114&116\\\hline m$_{H^{\pm}}$&272&230&263&256\\\hline 
m$_A$&260&215&251&243\\\hline m$_{\chi^0}$&92&73&72&78\\\hline 
m$_{\chi^0_2}$&159&125&126&135\\\hline
m$_{\chi^0_3}$&-245&-202&-237&-232\\\hline 
m$_{\chi^0_4}$&284&246&265&264\\\hline
m$_{\chi^{\pm}}$&157,284&120,-247&125,266&133,265\\\hline 
m$_{\tilde{\tau}_{1,2}}$&59,162&61,135&60,166&55,176\\\hline 
m$_{\tilde{e}_{1,2}}$&81,154&71,131&83,157&86,165\\\hline
m$_{\tilde{\nu}}$&132&104&135&144\\\hline  m$_{\tilde {\rm
t}_{1,2}}$&466,557&393,479&453,529&501,567\\\hline  m$_{\tilde {\rm
b}_{1,2}}$&499,523&419,436&475,501&515,544\\\hline  m$_{\tilde {\rm
u}_{1,2}}$&516,531&429,441&493,510&535,553\\\hline  m$_{\tilde {\rm
d}_{1,2}}$&517,537&430,448&494,516&535,558\\\hline  m$_{\tilde
g}$&640&533&498&538\\\hline
$\mu$&-236&-192&-226&-222\\\hline
$\sigma_{p\bar{p}\rightarrow \chi^+\chi^-}$&94.73& 295.92&296.22&214.04\\
(fb)&(120.75)&(356.12)&(358.92)&(262.56)\\\hline
$\sigma_{p\bar{p}\rightarrow \chi^{\pm}\chi^0_2}$&113.42&352.46&394.38&277.78\\
(fb)&(150.30)&(437.53)&(492.35)&(351.80)\\\hline 
\end{tabular}
\end{center}
\newpage
\begin{center}  Table 3 \end{center}
\begin{center}
\begin{tabular}{|c|c|c|c|c|c|}  \hline  &Final states&Scenario 1&Scenario
2&Scenario 3&Scenario 4 \\\cline{2-6} &2$\tau$ &0.406 & 0.388 & 0.350 & 0.349
\\\cline{2-6} &4$\tau$ &0.166 & 0.168 & 0.171 & 0.166 \\\cline{2-6} &6$\tau$
&0.017 & 0.018 & 0.021 & 0.019 \\\cline{2-6}
$p{\bar p}\rightarrow $&e($\mu)$3$\tau$ &0.148 & 0.151 & 0.156 & 0.158
\\\cline{2-6}
$\chi^+\chi^-$&e($\mu)$5$\tau$ &0.030 & 0.033 & 0.038 & 0.038 \\\cline{2-6}
&2e($2\mu)$4$\tau$ &0.013 & 0.015 & 0.017 & 0.018 \\\cline{2-6} 
&e$\mu$4$\tau$&0.027 & 0.029 & 0.035& 0.036 \\\cline{2-6}
&$\sigma_{(\chi^{+}\chi^-)}.B_{(\tau^+\tau^-)}$(fb)&300.56 &173.89  &116.35 &
428.90
\\ &$\sigma_{(\chi^{+}\chi^-)}.B_{(\tau^{+}\tau^{+}+\tau^{-}\tau^{-})}$&126.86 &
76.65 &56.06 &206.29 
\\\hline &3$\tau$ &0.549 & 0.534 & 0.494 & 0.327\\\cline{2-6} &5$\tau$ &0.112 &
0.116 & 0.121 & 0.078 \\\cline{2-6}
$p{\bar p}\rightarrow $&2e($2\mu$)3$\tau$ &0.044 & 0.045 & 0.049 &
0.132\\\cline{2-6}
$\chi^{\pm}\chi^0_2$&2e($2\mu$)5$\tau$ &0.009 & 0.009 & 0.012 &
0.031\\\cline{2-6} &e($\mu$)4$\tau$ &0.099 & 0.104 & 0.110 & 0.074\\\cline{2-6}
&3e(3$\mu$)4$\tau$ &0.008 & 0.009 & 0.011 & 0.029 \\\cline{2-6} &e2$\mu$4$\tau$
&0.008 & 0.009 & 0.011 & 0.029\\\cline{2-6}
 &$\sigma_{(\chi^{\pm}\chi^0_2)}.B_{3 \tau}$(fb)&456.83 &216.96  &167.76 &276.82 
\\ &$\sigma_{(\chi^{\pm}\chi^0_2)}.B_{(\tau^+\tau^-)}$&45.32 &43.74  &23.04
&233.75 
\\
&$\sigma_{(\chi^{\pm}\chi^0_2)}.B_{(\tau^{+}\tau^{+}+\tau^{-}\tau^{-})}$&45.32
&43.74 &23.04 & 233.75
\\\hline
\end{tabular}
\end{center}

\newpage
\begin{figure}[htb]
\centerline{ \DESepsf(prod.epsf width 14 cm) } \smallskip
\nonumber
\end{figure}
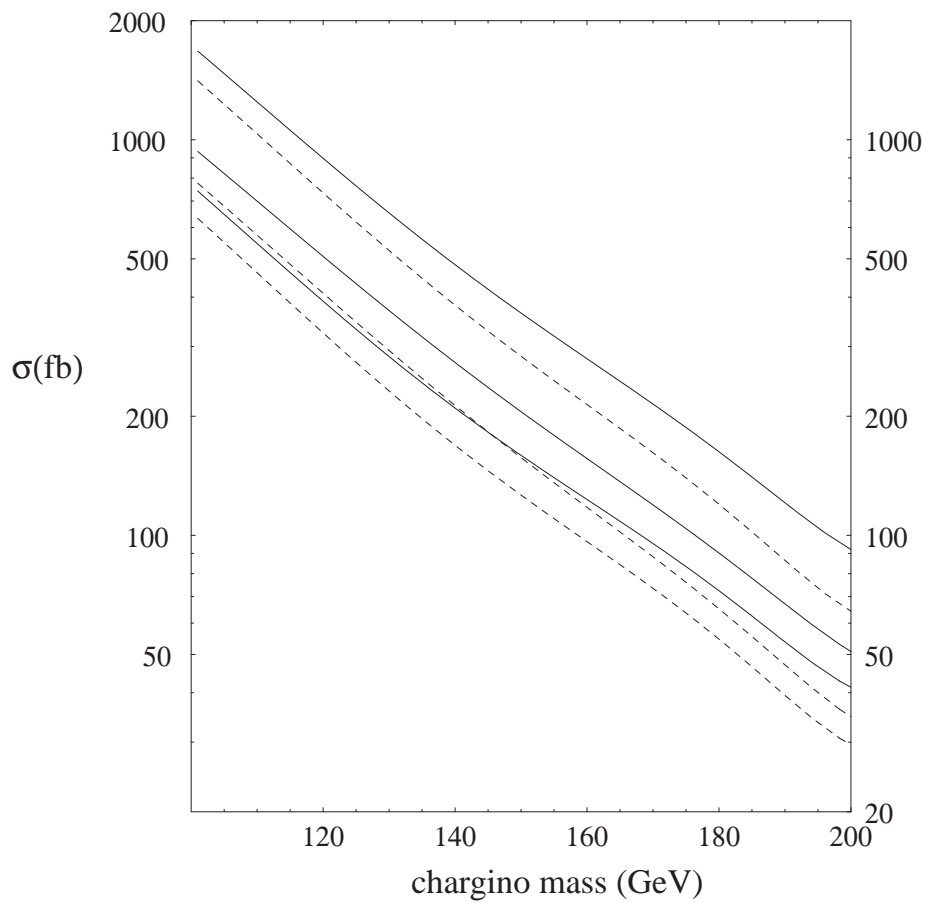
\end{document}